# PREVISÃO DOS PREÇOS DE ABERTURA, MÍNIMA E MÁXIMA DE ÍNDICES DE MERCADOS FINANCEIROS USANDO A ASSOCIAÇÃO DE REDES NEURAIS LSTM

**Gabriel de Oliveira Guedes Nogueira**
Instituto de Ciências Matemáticas e de Computação
Universidade de São Paulo
gabriel.gnogueira@usp.br

**Marcel Otoboni de Lima**
Instituto de Ciências Matemáticas e de Computação
Universidade de São Paulo
marcel.lima@usp.br

**Abstract**

In order to make good investment decisions, it is vitally important for an investor to know how to make good analysis of financial time series. Within this context, studies on the forecast of the values and trends of stock prices have become more relevant. Currently, there are different approaches to dealing with the task. The two main ones are the historical analysis of stock prices and technical indicators and the analysis of sentiments in news, blogs and tweets about the market. Some of the most used statistical and artificial intelligence techniques are genetic algorithms, Support Vector Machines (SVM) and different architectures of artificial neural networks. This work proposes the improvement of a model based on the association of three distinct LSTM neural networks, each acting in parallel to predict the opening, minimum and maximum prices of stock exchange indices on the day following the analysis. The dataset is composed of historical data from more than 10 indices from the world's largest stock exchanges. The results demonstrate that the model is able to predict trends and stock prices with reasonable accuracy.

**Resumo**

Para tomar boas decisões de investimento, é de vital importância para um investidor saber fazer boas *análises de séries temporais financeiras*. Nesse contexto, estudos sobre a previsão dos valores e tendências dos preços de ações listadas em bolsas de valores têm ganhado cada vez mais destaque. Atualmente, há diferentes abordagens para lidar com a tarefa. As duas principais são a análise histórica de preços e indicadores técnicos das ações e a análise de sentimentos em notícias, *blogs* e *tweets* sobre o mercado. Dentre as técnicas estatísticas e de inteligência artificial mais utilizadas, encontram-se algoritmos genéticos, *Support Vector Machines* (SVM) e diferentes arquiteturas de redes neurais artificiais. Este trabalho propõe a expansão de um modelo baseado na associação de três redes neurais LSTM distintas, cada qual atuando em paralelo para prever os preços de abertura, mínima e máxima de índices de bolsas de valores no dia seguinte ao analisado. O conjunto de dados é composto por dados históricos de mais de 10 índices das maiores bolsas de valores do mundo. Os resultados demonstram que o modelo não só é capaz de prever os preços com relativa acurácia, como também é capaz de prever sua tendência de oscilação.

## 1 Introdução

A análise de *séries temporais financeiras* é uma parte importante do processo de tomada de decisão de um investidor. O aumento e a queda dos preços das ações são influenciados por muitos fatores, tais como política, economia e sociedade. Para fazer boas escolhas de investimento, é preciso entender as tendências atuais do mercado e fazer previsões com base nelas. No contexto do mercado de ações, a aquisição de lucro está diretamente atrelada à acurácia das previsões feitas.

Para empresas listadas em bolsas de valores, o preço de suas ações não só reflete as condições operacionais da empresa e as expectativas de desenvolvimento futuro, como também é um índice técnico importante para suas pesquisas e tomadas de decisão. Ademais, a pesquisa envolvendo a previsão do preço de ações também tem um papel importante na análise do desenvolvimento econômico futuro de um país. Esse campo de estudo, portanto, tem grande significado teórico e amplas perspectivas de aplicação.

A hipótese do passeio aleatório [1] afirma que os preços das ações no mercado são definidos de forma aleatória e que, por tal motivo, é impossível prevê-los. Os avanços no campo da inteligência artificial e o aumento da quantidade de dados disponíveis, contudo, tornaram possível prever o comportamento do preço de ações com um desempenho melhor do que o de um processo aleatório [2, 3, 6-9, 12-16].

O presente trabalho tem por objetivo desenvolver um modelo de rede neural profunda capaz de prever, simultaneamente, o preço de abertura, o menor preço e o maior preço do dia seguinte de ações e índices de diferentes bolsas de valores. Para realizar as previsões, o modelo recebe como entrada séries temporais contendo os preços, volumes de negociação e alguns indicadores técnicos do ativo nos últimos dias. É proposto um modelo composto pela associação de múltiplas redes recorrentes do tipo *Long-Short Term Memory* (LSTM), que atuam com certo grau de independência para prever cada um dos três preços. Trata-se, como veremos, de uma expansão da ideia desenvolvida por Ding & Chin [2]. O desempenho do modelo foi comparado com os resultados obtidos por Ding & Chin [2].

O restante deste artigo é dividido em quatro seções principais. A seção **2** apresenta o estado atual da pesquisa sobre previsão do preço de ações. A seção **3** descreve os dados utilizados e como eles foram obtidos e explica a arquitetura do modelo proposto. A seção **4** apresenta os parâmetros experimentais e os resultados obtidos. Por fim, a seção **5** conclui o artigo, discutindo perspectivas para trabalhos futuros. Todo o código utilizado nos experimentos deste trabalho está disponível de forma aberta no GitHub[1].

---

[1] https://github.com/Talendar/stocks_prices_prediction

## 2 Trabalhos relacionados

Muitas pesquisas abordam a previsão dos preços e das tendências dos preços de ações listadas em bolsas de valores. Como indicado por Vachhani et al. [11], há três categorias principais de abordagens. A primeira, chamada de abordagem preditiva, usa preços históricos e indicadores técnicos como atributos de entrada para os modelos. A segunda, por sua vez, faz uso da análise de sentimentos em notícias e *blogs* sobre o mercado para prever o seu comportamento. Por fim, a terceira abordagem, chamada de híbrida, une as duas primeiras.

Kim [13] e Xia et al. [12] fizeram uso do algoritmo *Support Vector Machine* (SVM) para construir modelos de regressão com base em dados históricos para prever a tendência de ações. Essa abordagem é posteriormente otimizada por meio do uso do algoritmo de otimização por enxame de partículas, que aumentou consideravelmente a robustez do modelo baseado em SVM, acarretando, contudo, um aumento significativo nos custos computacionais envolvidos [14]. Outra abordagem frequentemente utilizada é a que faz uso do Modelo Auto-Regressivo Integrado de Médias Móveis [3], que, apesar de ser um conceito antigo, mostrou-se eficiente na previsão de séries temporais financeiras.

Os avanços recentes no campo de *deep learning*, tal como o aumento do poder computacional e da quantidade de dados disponíveis, tornou possível o uso de modelos mais avançados para a previsão de preços de ações. Dentre eles, destacam-se as redes LSTM. Jia [16] discutiu a efetividade desses tipos de rede para a previsão do preço de ações, demonstrando que se trata de um método efetivo para a tarefa. De fato, a classe de redes neurais mais utilizada para a previsão de séries temporais financeiras é, sem dúvida, a de redes recorrentes do tipo LSTM [2, 6, 7, 8, 9 e 15]. Selvin et al. [8] obtiveram bons resultados associando camadas LSTM à camadas convolucionais. Li et al. [9] propôs a adição de mecanismos de atenção em uma rede LSTM, o que levou a um aumento de desempenho.

Um modelo de alto-falante com múltiplas saídas baseado em redes RNN-LSTM foi usado no campo de síntese e adaptação de fala [17]. Os resultados experimentais mostram que o modelo é melhor do que o modelo de um único alto-falante. Um modelo de rede neural convolucional de múltiplas entradas e múltiplas saídas (MIMO-Net) foi projetado para segmentar células de imagens de microscópio de fluorescência [18]. Os resultados experimentais mostram que este método é superior ao método do estado da arte de segmentação baseado em *deep learning*. Fundamentando-se nessas duas pesquisas, Ding & Chin [2] propuseram um modelo que associa redes neurais LSTM em paralelo para prever, simultaneamente, o preço de abertura, o menor preço e o maior preço de ações no dia seguinte. Esta pesquisa foi inspirada por essa ideia, que defende que alguns parâmetros e indicadores de uma ação estão associados uns aos outros e devem, portanto, ser calculados de forma conjunta. Propomos, como será discutido adiante, uma expansão do modelo apresentado por Ding & Chin [2].

## 3 Material e métodos

### 3.1 Obtenção dos dados

Um índice de mercado é uma carteira hipotética de participações de investimento representativas de um segmento do mercado financeiro [4]. O cálculo do valor do índice provém dos preços das participações subjacentes. Neste trabalho, deu-se preferência ao uso de séries temporais financeiras envolvendo índices de mercado ao invés de ações de empresas individuais, haja vista a maior quantidade de dados históricos disponíveis para os índices. Ademais, por terem diversas empresas (muitas vezes de setores econômicos distintos) em seus portfólios, os índices de mercado são mais abrangentes, sendo bons indicativos da situação econômica geral do segmento financeiro que representam.

Os dados históricos de negociação dos índices considerados foram extraídos do site *Yahoo! Finance*[2] por meio de uma biblioteca *open-source* para a linguagem de programação *Python* chamada *Yahooquery*. Os indicadores técnicos, descritos em mais detalhes na próxima subseção, foram calculados de forma programática a partir de implementações feitas pelos autores. A tabela 1 lista os índices de mercado usados nos experimentos.

**Tabela 1.** Índices de mercado considerados.

| Ticker | Nome | País |
|---|---|---|
| BVSP | Índice BOVESPA | BR |
| NYA | NYSE Composite | US |
| DJI | Dow Jones Industrial Average | US |
| IXIC | NASDAQ Composite | US |
| RUT | Russell 2000 Index | US |
| FCHI | CAC 40 | FR |
| FTSE | Financial Times Stock Exchange 100 Index | EN |
| 000001.SS | SSE Composite Index | CH |
| HSI | Hang Seng Index | CH |
| N225 | Nikkei Stock Average | JP |
| BSESN | S&P Bombay Stock Exchange Sensitive Index | IN |

### 3.2 Seleção de *features*

Como atributos de entrada para os modelos, são usados os preços e volumes de negociação do índice, tal como indicadores técnicos calculados com base neles, dos últimos *n* dias. Como será visto na seção 4, diferentes valores para *n* foram testados nos experimentos.

**Tabela 2.** Atributos de entrada dos modelos.

| Preço de abertura | SMA5 | EMA10 |
|---|---|---|
| Preço de fechamento | SMA10 | EMA20 |
| Menor preço do dia | SMA20 | EMA40 |
| Maior preço do dia | SMA40 | MACD |
| Volume de negociação | EMA5 | |

Os indicadores técnicos utilizados foram a Média Móvel Simples (SMA), a Média Móvel Exponencial (EMA) e a Média Móvel

---
[2] https://finance.yahoo.com/

Convergência/Divergência (MACD), tendo sido considerados diferentes intervalos de tempo. A tabela 2 lista todas as *features* para cada um dos dias da sequência temporal de um exemplo de entrada.

### 3.3 Pré-processamento

Com o objetivo de trazer o preço dos índices para uma mesma escala de medida e de otimizar o treinamento dos modelos, os dados foram normalizados com a técnica de *min-max normalization*. Tal método redimensiona os valores do conjunto de treinamento para o intervalo [0, 1]. A normalização dos conjuntos de validação e de teste foi feita usando-se os parâmetros usados na normalização do conjunto de treinamento. A equação 1 apresenta a fórmula utilizada para normalizar os dados.

$$x_{norm} = \frac{x - min(x)}{max(x) - min(x)} \quad (1)$$

### 3.4 Rede neural LSTM

Redes neurais recorrentes (RNN) são uma classe de redes neurais artificiais especializadas no processamento de dados sequenciais. O principal ponto que diferencia uma RNN das redes tradicionais é a presença de *loops* em sua estrutura. Em princípio, RNNs podem utilizar tais conexões de *feedback* para armazenar informações recentes da sequência de entrada na forma de ativações [5]. Dessa forma, para fazer uma previsão, tal tipo de rede leva em consideração tanto a entrada do passo atual quanto informações vistas previamente na sequência que está sendo processada.

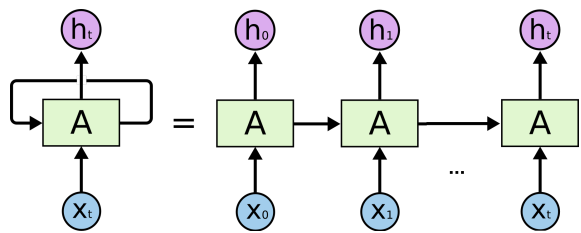

**Figura 1.** Estrutura geral de uma RNN.

Neste trabalho, foi usada uma arquitetura especial de RNNs conhecida como rede *Long-Short Term Memory* (LSTM), otimizada para lidar com dependências de longo prazo na sequência de entrada [5]. O uso desse tipo de rede para a previsão dos preços e das tendências de ativos financeiros têm demonstrado bons resultados [2, 6, 7, 8 e 9].

Para melhor lidar com dependências de longo prazo, redes LSTM possuem quatro componentes principais: uma *célula de memória*, um *portão de entrada*, um *portão de saída* e um *portão de esquecimento*. A *célula* é responsável por lembrar da dependência dos elementos da sequência de entrada por períodos arbitrariamente longos de tempo, enquanto que os portões controlam o fluxo de informação para dentro e para fora da *célula*.

Cada portão é composto por uma matriz de pesos, um termo *bias* e uma função de ativação sigmóide, cuja fórmula pode ser vista na equação 2. O vetor de saída de um portão possui valores entre 0 e 1, que podem ser interpretados como sendo a porcentagem da informação armazenada em cada componente correspondente de um vetor de referência que deverá ser mantida.

$$\sigma(x) = \frac{1}{1 + e^{-x}} \quad (2)$$

O *portão de esquecimento* determina quanto da informação presente na célula será mantida. Sua fórmula pode ser vista na equação 3, onde $f_t$ é a saída do portão no passo $t$, $W_f$ é a sua matriz de pesos, $x_t$ é a entrada da rede no passo $t$, $h_{t-1}$ é a saída da rede no passo $t - 1$ e $b_f$ é o termo *bias* do portão. Note que a entrada do portão é uma concatenação dos vetores $x_t$ e $h_{t-1}$, o que implica que informações previamente processadas pela rede estão sendo levadas em consideração juntamente com a entrada atual. Isso é também observado no restante da rede.

$$f_t = \sigma(W_f[x_t, h_{t-1}] + b_f) \quad (3)$$

O *portão de entrada*, por sua vez, controla quanto das novas informações obtidas serão adicionadas à *célula*. Sua fórmula é apresentada na equação 4, seguindo uma notação similar à usada anteriormente. A variável $i_t$ representa a saída do *portão de entrada* no passo $t$.

$$i_t = \sigma(W_i[x_t, h_{t-1}] + b_i) \quad (4)$$

O *portão de saída* computa quanto dos valores armazenados na *célula* serão usados para calcular a saída da rede. Sua fórmula pode ser vista na equação 5. A variável $o_t$ representa a saída do *portão de saída* no passo *t*.

$$o_t = \sigma(W_o[x_t, h_{t-1}] + b_o) \quad (5)$$

O valor $c_t$ assumido pela *célula* no passo *t* é determinado pelas saídas dos portões de esquecimento e de entrada, tal como pelo cálculo de um valor candidato $\hat{c}_t$ para a *célula*. As equações 6 e 7 apresentam as computações envolvidas nesta etapa.

$$\hat{c}_t = \tanh(W_c[x_t, h_{t-1}] + b_c) \quad (6)$$

$$c_t = f_t \odot c_{t-1} + i_t \odot \hat{c}_t \quad (7)$$

Por fim, a saída $h_t$ da rede ao fim do passo *t* é computada com base no novo valor da *célula* e no *portão de output*. Essa operação é representada pela equação 8.

$$h_t = o_t \odot \tanh c_t \quad (8)$$

**Figura 2.** Etapas do processamento em uma célula de uma rede LSTM.

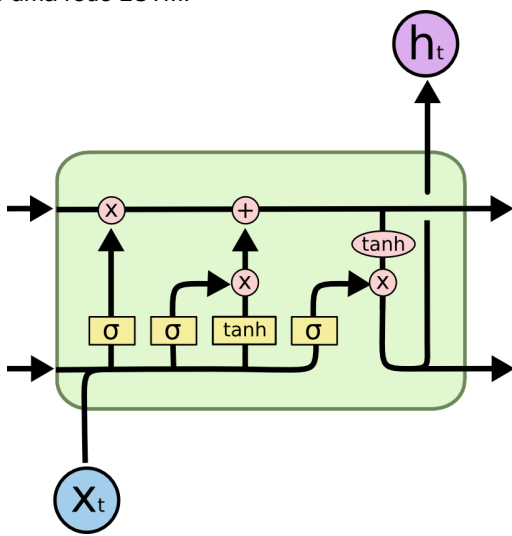

### 3.5 Associação de redes LSTM

Este trabalho propõe uma expansão do modelo apresentado por Ding & Chin [2]. Tal modelo é composto por uma associação de três redes LSTM paralelas, cada uma atuando com certo grau de independência. Inicialmente, os dados de entrada são processados por uma das redes, que irá prever o preço de abertura do dia seguinte. Em seguida, uma outra rede computa, com base na concatenação de um pré-processamento dos dados de entrada por uma camada LSTM única e do preço de abertura obtido, o preço mínimo do ativo no dia seguinte. Por fim, uma terceira rede computa, com base na concatenação do pré-processamento dos dados de entrada por uma outra camada LSTM única e dos preços de abertura e mínimo calculados, o preço máximo do ativo no dia seguinte.

**Figura 3.** Arquitetura do modelo apresentado por Ding & Chin [2], ao qual propomos uma expansão.

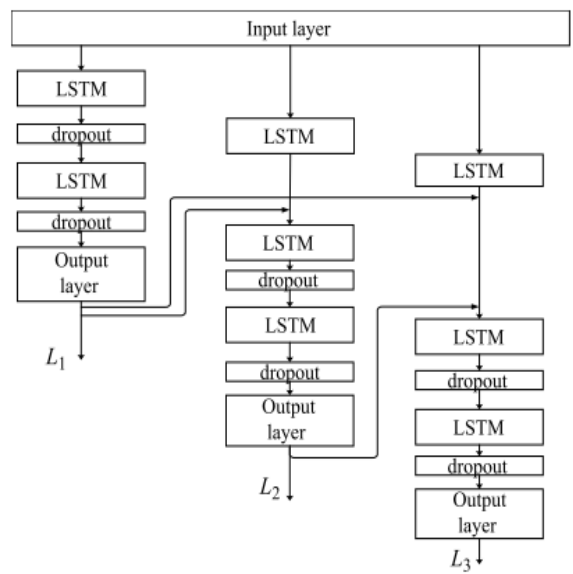

Na figura 3 é possível visualizar a arquitetura geral do modelo. Ao modelo proposto por Ding & Chin [2], acrescentamos camadas totalmente conectadas e aumentamos o número de camadas LSTM ocultas. Ademais, adaptamos o modelo para que ele fosse capaz de receber como entrada, além dos preços históricos e volumes de negociação, os indicadores técnicos listados anteriormente na tabela 2. Ao invés das camadas de *dropout* utilizadas por Ding & Chin [2], optamos por utilizar o *recurrent* *dropout*, uma forma de regularização otimizada para não causar perdas de memórias de longo prazo [10].

### 3.6 Métodos de avaliação

Para possibilitar a avaliação da capacidade de generalização dos modelos, utilizou-se a técnica de validação cruzada. O conjunto de dados, nesse contexto, foi dividido em três subconjuntos: um conjunto de treinamento, com 85% dos dados, um conjunto de validação, com 10% dos dados, e um conjunto de testes, com 5% dos dados. O conjunto de treinamento contém os dados históricos dos índices de 01/01/2007 até 19/10/2018. O conjunto de validação, por sua vez, engloba os pregões de 22/10/2018 até 18/03/2020. Por fim, o conjunto de dados possui o histórico entre as datas 19/03/2020 e 10/12/2020. No total, há 38209 pregões, considerando todos os índices.

Como métricas de avaliação do desempenho dos modelos, usou-se o erro quadrático médio (MSE), o erro absoluto médio (MAE) e o erro percentual absoluto médio (MAPE). Ademais, consideramos, ainda, a acurácia de acertos dos modelos das tendências dos preços. Considera-se que o modelo previu corretamente a tendência do preço quando ele acerta o sentido de oscilação do preço (variação positiva ou negativa).

## 4 Experimentos

Para treinar o modelo, foi utilizado o algoritmo *backpropagation through-time*, em associação ao otimizador *Adaptive Moment Estimation* (ADAM). Foram realizados experimentos usando-se diferentes janelas de intervalo da série temporal (quantos dias anteriores a rede deve considerar ao fazer uma previsão). Os melhores resultados foram obtidos alimentando a rede com dados referentes aos últimos 7 pregões do mercado. O modelo foi treinado com uma taxa de aprendizado de $6 . 10^{-4}$ por 100 épocas. Na figura 4, encontra-se o gráfico com a evolução da função de perda no conjunto de treinamento e no conjunto de validação ao longo da sessão de treinamento. É possível observar a alta oscilação do erro no conjunto de validação ao longo do treino, fator provavelmente derivado do fato de serem usados diferentes índices de ações no mesmo conjunto de dados.

**Figura 4.** Erro quadrático médio (MSE) no eixo y ao longo de 100 épocas no eixo x.

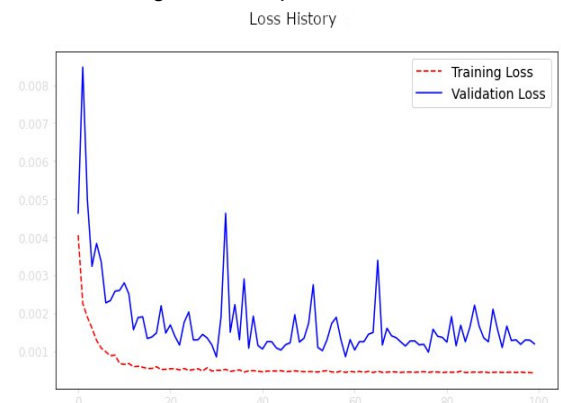

Após o treinamento, durante o qual são feitos *checkpoints* dos modelos, escolhemos o modelo que apresentou o menor erro (MSE) no conjunto de validação como modelo final. Na tabela 3, mostramos os erros quadráticos médios obtidos pelo nosso modelo em cada um dos conjuntos de dados e comparamos com os resultados obtidos por Ding & Chin [2], considerando 100 épocas de treinamento. Apesar de Ding & Chin [2] usarem o mesmo método de normalização que nós (*min-max normalization*), de forma que seus dados encontram-se na mesma escala de medida que os nossos, eles utilizam ações e índices diferentes daqueles que utilizamos.

**Tabela 3.** Erro quadrático médio (MSE) em diferentes conjuntos de dados.

| Modelo | Erro de treino | Erro de validação | Erro de teste |
|---|---|---|---|
| Nosso | $1,5 . 10^{-4}$ | $8,4 . 10^{-4}$ | $1,7 . 10^{-3}$ |
| Ding & Chin [2] | - | - | $2,9 . 10^{-2}$ |

Como pode-se observar, a expansão do tamanho das redes do modelo, tal como o uso do *recurrent dropout* e de mais dados de treinamento, acarretou uma melhora significativa no desempenho do nosso modelo em relação ao de Ding & Chin [2]. Na tabela 5, pode-se conferir os valores para o erro percentual absoluto médio (MAPE) para cada um dos índices considerados nos dados utilizados.

**Tabela 4.** Erro percentual absoluto médio geral e para cada um dos tipos de preço.

| Ticker | MAPE geral (%) | MAPE open (%) | MAPE low (%) | MAPE high (%) |
|---|---|---|---|---|
| FTSE | **0.85** | **0.42** | 0.93 | 1.10 |
| FCHI | 1.21 | 0.87 | 1.34 | 1.47 |
| HSI | 1.12 | 0.94 | 1.78 | **0.81** |
| 000001.SS | 0.92 | 0.50 | **0.77** | 1.48 |
| N225 | 0.88 | 0.68 | 0.95 | 1.00 |
| BSESN | 1.36 | 0.92 | 1.09 | 2.06 |
| IXIC | 6.34 | 4.84 | 5.83 | 8.35 |
| BVSP | 1.59 | 0.89 | 1.47 | 2.42 |
| RUT | 1.15 | 0.63 | 1.34 | 1.48 |
| NYA | 0.96 | 0.84 | 1.05 | 0.97 |
| DJI | 1.05 | 0.68 | 1.03 | 1.45 |

A análise da tabela 4 nos permite concluir que o modelo teve mais facilidade em prever os preços de alguns índices em relação a outros. O modelo foi capaz, por exemplo, de prever razoavelmente bem os preços do índice FTSE, referente à bolsa de valores de Londres, mas apresentou dificuldade para prever os preços do IXIC, referente à bolsa de valores americana NASDAQ. Na figura 5, pode-se observar o gráfico comparando os preços reais do índice FTSE com os preços previstos pelo modelo.

**Figura 5.** Gráfico com a oscilação real dos preços (em verde) e com os preços previstos pelo modelo (laranja) para o índice FTSE no conjunto de testes.

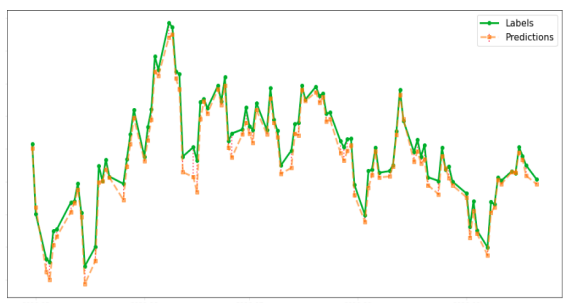

Na tabela 5 é possível ver a acurácia das previsões da tendência de oscilação (positiva ou negativa) do preço do índice pelo modelo.

**Tabela 5.** Acurácia de tendências prevista pela rede.

| Ticker | Tend. open (%) | Tend. low (%) | Tend. high (%) |
|---|---|---|---|
| FTSE | **85** | 58 | 52 |
| FCHI | 73 | 55 | 52 |
| HSI | 67 | 42 | 62 |
| 000001.SS | 81 | 63 | **63** |
| N225 | 64 | 55 | 53 |
| BSESN | 75 | 55 | 37 |
| IXIC | 46 | 44 | 35 |
| BVSP | 80 | 63 | 54 |
| RUT | 84 | **64** | 46 |
| NYA | 73 | 59 | 63 |
| DJI | 83 | 63 | 51 |

A tabela 5 indica que o modelo aprendeu muito bem a prever os preços de abertura, obtendo valores acima de 80% na previsão da tendência dos preços para algumas ações. Porém, para as previsões de preços mínimo e máximo, o modelo não foi capaz de aprender muito além do que uma previsão aleatória seria capaz, obtendo resultados que flutuam em torno do 50% e nada muito além dos 60%.

## 5 Conclusão

Neste trabalho, foi proposta uma expansão para o modelo de redes neurais LSTM associadas proposto por Ding & Chin [2] para prever os preços de abertura, mínima e máximo de ações. Nosso modelo foi capaz de alcançar um desempenho melhor do que aquele no qual foi inspirado, obtendo resultados promissores. Isso indica que a adição de mais camadas profundas nas redes, tal como a utilização do *recurrent dropout* e de mais dados foi benéfica ao modelo. Para trabalhos futuros, buscaremos testar novos indicadores técnicos como entrada e adicionar mecanismos de atenção nas redes LSTM.

# 6 Referências